\documentclass{IEEEtran4PSCC}
\usepackage{xcolor}

\ifCLASSINFOpdf
   \usepackage[pdftex]{graphicx}
\else
   \usepackage[dvips]{graphicx}
\fi
\usepackage[cmex10]{amsmath}
\usepackage{arydshln}
\usepackage{tabularx}
\usepackage{booktabs}
\usepackage{makecell}
\usepackage{cite}
\usepackage{amsmath,amssymb}
\usepackage{enumitem}
\usepackage{algorithm}
\usepackage{float}
\usepackage{algpseudocode}
\usepackage{bm}
\usepackage{longtable}
\usepackage{color}
\usepackage{url}
\usepackage{multicol}
\usepackage{comment}
\usepackage{multirow}
\usepackage{graphicx} 
\usepackage{mathtools} 
\allowdisplaybreaks    
\usepackage{tikz}
\usepackage{amsfonts} 
\usepackage{setspace}
\usepackage{hyperref}
\usepackage{enumitem}
\usepackage{array}

\makeatletter
\let\old@ps@headings\ps@headings
\let\old@ps@IEEEtitlepagestyle\ps@IEEEtitlepagestyle
\def\psccfooter#1{%
    \def\ps@headings{%
        \old@ps@headings%
        \def\@oddfoot{\strut\hfill#1\hfill\strut}%
        \def\@evenfoot{\strut\hfill#1\hfill\strut}%
    }%
    \def\ps@IEEEtitlepagestyle{%
        \old@ps@IEEEtitlepagestyle%
        \def\@oddfoot{\strut\hfill#1\hfill\strut}%
        \def\@evenfoot{\strut\hfill#1\hfill\strut}%
    }%
    \ps@headings%
}
\makeatother

\begin{document}
%
\title{Quantifying the Impact of Missing Risk Markets for Decarbonized Power Systems with Long Duration Energy Storage}

\author{
\IEEEauthorblockN{Andreas C. Makrides$^1$, Adam Suski$^2$, Elina Spyrou$^2$ }
\IEEEauthorblockA{$^1$Dyson School of Design Engineering, Imperial College London, London, United Kingdom \\
$^2$ Department of Electrical and Electronic Engineering, Imperial College London, London, United Kingdom \\
\{andreas.makrides24@imperial.ac.uk, a.suski23@imperial.ac.uk, evangelia.spyrou@imperial.ac.uk\}}
}


\maketitle

\begin{abstract}
The transition to a fully decarbonized electricity system depends on integrating new technologies that ensure reliability alongside sustainability. However, missing risk markets hinder investment in reliability-enhancing technologies by exposing investors to revenue uncertainty. This study provides the first quantitative assessment of how missing risk markets affect investment decisions in power systems that depend on long-duration energy storage (LDES) for reliability. We develop a two-stage stochastic equilibrium model with risk-averse market participants, which independently sizes power and energy capacity. We apply the method to a case study of a deeply decarbonized power system in Great Britain. The results show that incomplete risk markets reduce social welfare, harm reliability, and discourage investment in LDES and other technologies with volatile revenue streams. Revenue volatility leads to substantial risk premiums and higher financing costs for LDES, creating a barrier to its large-scale deployment. These findings demonstrate the importance of policy mechanisms that hedge revenue risk to lower the cost of capital and accelerate investment in reliability-enhancing, zero-carbon technologies.

\end{abstract}
\begin{IEEEkeywords}
long-duration energy storage, market (in)completeness, power system planning, risk management.
\end{IEEEkeywords}

\vspace{0.35em}
\noindent\textit{Sets}
\vspace{0.35em}

\begin{tabular}{@{\hspace{-0.7em}}l@{\hspace{4.5em}}p{0.75\columnwidth}}
$\mathcal{A}$         & Market participants ($\mathcal{A}=\mathcal{R}\cup\mathcal{C}$). \\
$\mathcal{G}$         & Generators. \\
$\mathcal{S}$         & Storage assets. \\
$\mathcal{R}$         & Resources ($\mathcal{R}=\mathcal{G}\cup\mathcal{S}$). \\
$\mathcal{C}$         & Consumers. \\
$\mathcal{T}$         & Time steps. \\
$\mathcal{O}$         & Scenarios. \\
\end{tabular}

\vspace{0.35em}
\noindent\textit{Decision Variables}
\vspace{0.35em}

\begin{tabular}{@{\hspace{-0.7em}}l@{\hspace{1em}}p{0.85\columnwidth}}
$x^{\text{P}}_r$  & Installed power capacity of resource $r$ [MW]. \\
$x_s^{\text{E}}$  & Installed energy capacity of storage $s$ [MWh]. \\
$q_{g,t,o}$  & Energy generation at time $t$ in scenario $o$ [MW]. \\
$q^{\text{ch}}_{s,t,o}, q^{\text{dch}}_{s,t,o}$  & Storage charging/discharging power [MW]. \\
$e_{s,t,o}$  & Storage state-of-charge [MWh]. \\
$d^{\text{fix}}_{t,o}$,$d^{\text{flex}}_{t,o}$  & Fixed/Flexible demand [MW].\\ 
$\zeta_a$  & Value-at-Risk (VaR) for participant [\$]. \\
$u_{a, o}$  & Participant surplus [\$]. \\
$u^+_{a, o}$  & Auxiliary CVaR variable for participant [\$/]. \\
$\rho_a$  & Participant risk-adjusted surplus [\$]. \\
$\theta_{a ,t, o}$  & Participant electricity production [MW]. \\
\end{tabular}

\vspace{0.35em}
\noindent\textit{Parameters}
\vspace{0.35em}

\begin{tabular}
{@{\hspace{-0.7em}}l@{\hspace{2.7em}}p{0.85\columnwidth}}
$W_{t,o}$  & Time weights [h]. \\
$A_{g,t,o}$  & Generators availability [\%]. \\
$P_o$  & Scenario probability [\%]. \\
\end{tabular}
\begin{tabular}
{@{\hspace{+0.4em}}l@{\hspace{1em}}p{0.78\columnwidth}}
$B$  & Value of lost load (VOLL) [\$/MWh]. \\
$\delta$  & Risk aversion coefficient. \\
$\Psi$  & CVaR confidence level. \\
$D^{\text{flex}}_{t, o}$,$D^{\text{fix}}_{t, o}$  & Elastic/Inelastic share of the demand [MW]. \\
$\lambda_{t,o}$  & Energy price [\$/MWh]. \\
$r_{t,o}$  &  System imbalance [MW]. \\
$C^{\text{VAR}}_r$  & Resource variable cost [\$/MWh]. \\
$C^{\text{INV,P}}_r$  & Annualised resource investment cost [\$/MW-yr]. \\
$C^{\text{INV,E}}_s$  & Annualised storage energy component investment cost [\$/MWh-yr]. \\
$\eta^{\text{ch}}_s$,$\eta^{\text{dis}}_s$  & Charging/Discharging efficiency [\%]. \\
$r_g$  & Generator ramp rate limit, \(\in\) [0,1]. \\
$\underline{m}$  & Minimum generation fraction, \(\in\) [0,1]. \\
$\overline{c}$  & Maximum generation capacity factor, \(\in\) [0,1]. 
\end{tabular}

\vspace{0.35em}
\noindent\textit{ADMM Hyper-Parameters}
\vspace{0.35em}

\begin{tabular}{@{\hspace{-0.7em}}l@{\hspace{4.3em}}p{0.85\columnwidth}}
$\gamma$  & Penalty factor. \\
$j$  &  ADMM's iteration $j^{th}$. \\
$J$  &  Maximum number of iterations for ADMM. \\
$\epsilon$  & Convergence parameter. \\
$\tau$  & Convergence tolerance.\\
\end{tabular}

\thanksto{\noindent This work was supported by the Energy Futures Lab, a Leverhulme International Professorship (LIP-2020-002) and the Engineering and Physical Sciences Research Council (EP/Y025946/1).}

\section{Introduction}

The integration of variable renewable energy (VRE) sources into power systems poses challenges for balancing supply and demand \cite{pauldenholm}. As power systems transition toward net-zero emissions, the temporal scope of balancing expands beyond hours and days to weeks and seasons \cite{InternationalEnergyAgencyIEA2023COP28Gap}. Large-scale integration of energy storage systems (ESS) can play a critical role in tackling these challenges \cite{InternationalEnergyAgencyIEA2011HarnessingChallenge}. Currently, most installed storage consists of short-duration lithium-ion batteries, complemented by pumped hydro, which remains the dominant technology for large-scale, long-duration storage globally \cite{Schmidt2023MonetizingStorage}. While batteries can economically and technically manage demand-supply imbalances over short timescales, they are not cost-effective for operations that extend beyond a few hours due to their relatively high energy-specific investment costs. Therefore, to further advance decarbonisation, it is essential to integrate storage technologies that can effectively store and release energy over longer durations\cite{DepartmentforBusiness2022BenefitsStorage}.

Long-duration energy storage (LDES) technologies\footnote{LDES typically refers to storage with durations exceeding eight to twelve hours, significantly longer than those of most installed batteries.}, such as hydrogen storage, could provide flexibility with durations long enough to support a reliable electricity supply in a deeply decarbonized power system. Investment in LDES ensures reliability even as conventional backup units retire, because LDES can store energy at times of VRE surplus and release energy during times of low VRE supply, particularly during so-called energy droughts\cite{Sepulveda2021TheSystems}. In the UK, LDES is becoming an important part of future decarbonization plans, with projections indicating that over 3 TWh of LDES will be needed in a net-zero energy system \cite{fes2025}. LDES solutions are particularly cost-effective when clean firm alternatives, such as nuclear power or carbon capture and storage (CCS), are limited or prohibitively expensive \cite{deSisternes2016TheSector, LDESCouncil2021Net-zeroGrid, govrep2024}.

While planning models find integration of LDES beneficial in net-zero systems \cite{afry2022}, in practice, LDES investments face several barriers. Existing markets do not necessarily provide efficient incentives for the dispatch of long-duration storage \cite{Bassi2015BridgingUnion, McNamara2022Long-durationOpportunities, FelicityJones2016CrackingThem, Bowen2019Grid-ScaleQuestions}. Moreover, these assets face financing challenges due to high capital costs and long construction times \cite{Sepulveda2021TheSystems}. The financing challenges are exacerbated by the lack of revenue certainty \cite{callforevidence}. LDES revenues depend on opportunities for long-term energy arbitrage, which are sensitive to VRE outputs and demand. Hence, inter-annual weather variability, which affects VRE generation and electricity demand, can result in high volatility of LDES revenues \cite{adamsmissingmoney}. In a risk-averse environment, this volatility translates into increased financial risk \cite{kittelvariability}, which can jeopardize the bankability of LDES investments. 

Hedging contracts can help investors manage financial risk. The availability, variety, and liquidity of these contracts can significantly affect investment in risky, yet profitable in expectation, assets. In practice, electricity markets are often to a great extent incomplete, i.e., they lack all contracts necessary for fully hedging financial risk \cite{Newbery2016MissingInterconnectors, Ehrenmann2011GenerationAnalysis, deMaeredAertrycke2017InvestmentContracts}. In incomplete markets, investment in assets like LDES is usually lower than required for cost-effectively achieving overarching decarbonization goals\cite{Dimanchev2024ConsequencesEmissions}. To address this issue and accelerate LDES deployment, some regulators, such as Ofgem in Great Britain, are developing state-backed contracts \cite{ofgemldes}.

While policies are being developed to address the lack of revenue certainty for LDES, the literature lacks evidence on the effects of incomplete markets on LDES investment. Furthermore, policy discussions rely on hypothetical examples when illustrating how the introduction of hedging instruments could affect the cost of capital for LDES projects \cite{cepa_cap_2025}.

This article addresses this gap and shows how missing hedging instruments affect LDES investments. To achieve this objective, similar to \cite{Mays2019AsymmetricMarkets,Mays2023FinancialPenetration,Shu2023BeyondObligations,Dimanchev2024ConsequencesEmissions}, we formulate an equilibrium model with risk-averse investors deciding the capacities for electricity generation and storage. We find solutions using the alternating direction method of multipliers (ADMM) \cite{Boyd2010DistributedMultipliers}. Analysing the solutions for a deeply decarbonised case of the Great Britain (GB) power system, we answer the following research questions:
\begin{enumerate}
    \item How does (in)completeness affect social welfare and outcomes for consumers?
    \item How does the (un)availability of hedging instruments affect investment in LDES technologies in deep decarbonisation scenarios in a risk-averse environment?  
    \item How does risk aversion influence financing costs in the absence of financial instruments for hedging?
\end{enumerate}

To answer these questions, we compare two extreme cases in terms of the availability of hedging contracts. The first case is that of complete markets, where market participants, including LDES, can fully hedge their risks with zero transaction costs. The second case is that of fully incomplete markets, where hedging contracts are not available and investors independently manage their risk exposure by adjusting their investment levels. 

The modelling approach in this manuscript is drawn from the literature in risk-averse generation capacity expansion  \cite{Mays2019AsymmetricMarkets, Kaminski2021ImpactCompanies, Dimanchev2024ConsequencesEmissions}. In contrast to the existing literature, we independently size the power and energy components on LDES assets. This flexibility enables us to examine how the lack of hedging contracts influences decisions regarding storage durations.

The rest of the paper is organized into five parts. Section II presents the methodology we follow to find investment equilibria in fully incomplete electricity markets. Section III describes the case study, and Section IV explains the results of this work. Finally, Section V discusses key findings on the effects of incomplete markets on investments in LDES assets.

\section{Methodology}

\subsection{Modelling Framework}
The capacity investment problem is formulated as a two-stage stochastic equilibrium with risk-averse market participants. Let $\mathcal{A}$ denote the set of all participants, comprising resources $\mathcal{R}$ and consumers $\mathcal{C}$, with $\mathcal{A}=\mathcal{R}\cup\mathcal{C}$. Resources are either generators $\mathcal{G}$ or storage operators $\mathcal{S}$, so that $\mathcal{R}=\mathcal{G}\cup\mathcal{S}$. In the first stage, resources invest in power and, for storage, also energy capacity. In the second stage, after the uncertainty is realized, all participants determine operational variables. The second-stage scenarios, indexed with $o \in \mathcal{O}$ and occurring with probability $P_o$, represent alternative realizations of demand and VRE availability.

Each participant makes decisions to maximize its individual risk-adjusted surplus $\rho_a$, subject to physical and market-clearing constraints. Risk preferences are captured through a coherent risk measure defined as a convex combination of the expected value and the Conditional Value-at-Risk (CVaR) at confidence level $\Psi$ \cite{TyrrellRockafellar1998OptimizationValue-at-risk}. The generic form of each partipant's maximization problem is as follows:
\begin{subequations} \label{eq:generic}
\renewcommand{\theequation}{\theparentequation\alph{equation}} 
\begin{alignat}{1}
& \max_{\Theta_a} \; \rho_{a}  = (1 - \delta_a)  
\left( \zeta_a - \frac{1}{\Psi_a} \sum_{o} P_{o}\, u^+_{a,o} \right) 
\notag \\ 
& \quad + \delta_a  \left( \sum_o P_o\, u_{a,o} \right) \label{eq:gen_obj} \\[1ex]
& \zeta_a - u_{a,o} \leq u^+_{a,o}, \quad \forall o \quad (\mu_{a,o}) \label{eq:gen_cvar1} \\
& 0 \leq u^+_{a,o}, \quad \forall o \label{eq:gen_cvar2} \\
& \Theta_a \in \mathcal{F}_a, \quad  \label{eq:gen_feas}
\end{alignat}
\end{subequations}

The objective function \eqref{eq:gen_obj} is a weighted average of the expected value and CVaR of each participant's surplus $u_{a,o}$. Following \cite{TyrrellRockafellar1998OptimizationValue-at-risk}, in constraints \eqref{eq:gen_cvar1}–\eqref{eq:gen_cvar2}, $\zeta_a$ and $u^+_{a,o}$ correspond to the Value-at-Risk  and auxiliary surplus variable in scenario $o$, respectively. 
$\Theta_a$ includes the decision variables for participant $a$ and $\mathcal{F}_a$ is the feasible region for these decisions. In the following subsections, we provide detailed formulations for $u_{a,o}$ and $\mathcal{F}_a$ for each type of participant. 

\subsubsection{Resources}
The surplus of each resource $r \in \mathcal{R}$ is defined as the difference between its revenues and costs:
\begin{align}
u_{r, o} = \pi_{r, o} - C_{r, o}, \quad \forall r, o \label{eq:costsres}
\end{align}

For generators, the feasible region is the intersection of the following constraint sets:
\begin{subequations} \label{eq:gen}
\renewcommand{\theequation}{\theparentequation\alph{equation}} 
\begin{alignat}{1}
& C_{g, o} = C^{\text{V}}_g \sum_t W_{t,o} q_{g,t,o} 
+ C^{\text{I,P}}_g x^{\text{P}}_g, \quad \forall g, o \label{eq:costsgenres} \\
& \pi_{g, o} = \sum_{t}  W_{t,o}\, \lambda_{t,o}\, q_{g,t,o}, \quad \forall g,o \label{eq:revgen} \\
& 0 \leq q_{g,t,o} \leq x^{\text{P}}_g  A_{g, t, o}, \quad \forall g, t, o \label{eq:availgen} \\
& q_{g,t,o} - q_{g,t-1,o} \leq r_g  x^P_{g}, \quad \forall g, t, o \label{eq:r1} \\
& q_{g,t-1,o} - q_{g,t,o} \leq r_g  x^P_{g}, \quad \forall g, t, o \label{eq:r2} \\
& \underline{m}  x^P_{g} \leq q_{g,t,o}, \quad \forall g, t, o \label{eq:r3} 
\end{alignat}
\end{subequations}

Equation \eqref{eq:costsgenres} defines generation costs, while \eqref{eq:revgen} specifies revenues as the product of market prices and output. Constraints \eqref{eq:availgen}–\eqref{eq:r3} impose availability, ramping, and minimum stable generation requirements. 

For storage units, the constraints of the feasible region are:
\begin{subequations} \label{eq:stor}
\renewcommand{\theequation}{\theparentequation\alph{equation}} 
\begin{alignat}{1}
& C_{s, o} = C^{\text{INV,P}}_{s} x^{\text{P}}_{s}  
+ C^{\text{INV,E}}_{s} x^{\text{E}}_{s}, \quad \forall s,o \label{eq:costsstorres}\\
& \pi_{s, o} = \sum_{t} W_{t,o}\, \lambda_{t,o}\,
\big(q^{\text{dch}}_{s,t,o} - q^{\text{ch}}_{s,t,o}\big), \quad \forall s,o \label{eq:revstor}\\
& 0 \leq e_{s,t,o} \leq x^{\text{E}}_s, \quad \forall s, t, o \quad (\kappa^e_{s,t,o}) \label{eq:stor1} \\
& 0 \leq q^{\text{ch}}_{s,t,o} \leq x^{\text{P}}_s, \quad \forall s, t, o  \quad (\kappa^{ch}_{s,t,o}) \label{eq:stor2}\\
& 0 \leq q^{\text{dch}}_{s,t,o} \leq x^{\text{P}}_s, \quad \forall s, t, o \quad (\kappa^{dch}_{s,t,o}) \label{eq:stor3}\\
& e_{s,t,o} = e_{s,t-1,o} -  W_{t,o}  
\left(\frac{q^{\text{dch}}_{s,t,o}}{\eta^{\text{dis}}_s} 
- \eta^{\text{ch}}_s\, q^{\text{ch}}_{s,t,o} \right), \; \forall s, t, o \label{eq:stor4} \\
& e_{s,T_{\text{end}},o} \geq \text{SOC}^{\text{init}}_s\, x^{\text{E}}_s, \quad \forall s,o \label{eq:stor5}
\end{alignat}
\end{subequations}

Constraints \eqref{eq:stor1}–\eqref{eq:stor5} enforce energy balance, charging and discharging limits, storage efficiency, and terminal state-of-charge conditions. 

\subsubsection{Consumers}
Each consumer $c \in \mathcal{C}$ defines its surplus as the difference between the benefit $H_{t,o}$ from electricity consumption and the payments for electricity:
\begin{align}
u_{c, o} =   \sum_t W_{t,o} \left( H_{t,o} -  \lambda_{t,o} \left( d^{\text{fix}}_{t,o} + d^{\text{flex}}_{t,o}\right)\right), \quad \forall o \label{eq:utility}
\end{align}

For consumers, the constraints for the feasible region are:
\begin{align}
H_{t,o} =   B  
\left( d^{\text{fix}}_{t,o} + d^{\text{flex}}_{t,o} 
- \frac{(d^{\text{flex}}_{t,o})^2}{2 D^{\text{flex}}_{t,o}} \right), \quad \forall t,o \label{eq:ho}
\end{align}
\begin{subequations} \label{eq:con}
\renewcommand{\theequation}{\theparentequation\alph{equation}} 
\begin{alignat}{1}
& 0 \leq d^{\text{fix}}_{t,o} \leq D^{\text{fix}}_{t,o}, \quad \forall t, o \label{eq:ED_demand_fix} \\
& 0 \leq d^{\text{flex}}_{t,o} \leq D^{\text{flex}}_{t,o}, \quad \forall t, o \label{eq:ED_demand_flex}
\end{alignat}
\end{subequations}

Equations \eqref{eq:ho} defines consumer utility, and constraints \eqref{eq:ED_demand_fix}–\eqref{eq:ED_demand_flex} bound fixed and flexible demand. 

\subsubsection{Market Clearing}
The market-clearing condition is defined as the energy balance in each time step, taking into account generation dispatch levels, storage operations, and total consumption. To simplify notation, we define $\theta_{a,t,o}$ as the production of each market participant. \footnote{For consumers, production refers to negative consumption.}
\begin{subequations} \label{eq:ED_balance}
\begin{align}
& \text{Generators}: \theta_{g,t,o} = q_{g,t,o}, \quad \forall g, t, o \label{eq:pengeng1}\\
& \text{Storage assets}: \theta_{s,t,o} = q^{\text{dch}}_{s,t,o} - q^{\text{ch}}_{s,t,o}, \quad \forall s, t, o \label{eq:pengens2}\\
& \text{Consumers}: \theta_{c,t,o} = - d^{\text{fix}}_{t,o} - d^{\text{flex}}_{t,o}, \quad \forall c, t, o \label{eq:pengenc3}\\
& \sum_{a} \theta_{a,t,o} = 0, \quad (\lambda_{t,o})
\end{align}
\end{subequations}

\subsection{Complete Markets} \label{SUB: CompleteMarketsForm}

When the markets are complete, the equilibrium capacity investment problem can be reformulated as a single social welfare maximization problem, which is equivalent to the problem of a risk-averse central planner \cite{ralphandsmeers}.  The problem has the form of \eqref{eq:generic}, with the following scenario-specific social surplus:
\begin{equation}
    u^{COM}_o = H_o - \sum_r C_{r, o} , \quad \forall o \label{eq:costs}
\end{equation} and constraints for market-clearing and all participants. The solution to this problem implicitly considers that agents can hedge themselves against tail events with Arrow–Debreu securities for each scenario  \cite{Arrow1954ExistenceEconomy}. 

\subsection{Fully Incomplete Markets}

Finding incomplete investment equilibria is inherently complex because the problem lacks a straightforward equivalent optimization form \cite{Abada2017OnApproach}. By deriving and concatenating the Karush-Kuhn-Tucker (KKTs) optimality conditions, such problems are commonly reformulated into mixed complementary problems (MCPs).  While PATH and other Newton-based solvers can efficiently handle small-scale MCPs \cite{Gabriel2013ComplementarityMarkets}, they face significant computational and scalability limitations when applied to larger systems. In this article, ADMM is utilised \cite{Boyd2010DistributedMultipliers}, which has been shown to effectively find investment equilibria in incomplete electricity markets \cite{Hoschle2018AnMarkets, Kaminski2021ImpactCompanies}. 

ADMM is an iterative algorithm. In this article, each iteration involves solving the individual optimization problems of all participants (\ref{eq:generic}), given updated market prices. Following \cite{Hoschle2018AnMarkets, Kaminski2021ImpactCompanies}, prices are iteratively adjusted based on the system imbalance ($r_{t,o}$). The system imbalance is defined as the difference between total supply and demand at each hour. A positive system imbalance indicates overproduction and, as a result, prices decrease  in the subsequent iteration $j+1$, while a negative system imbalance shows underproduction and prices increase in the subsequent iteration $j+1$:
\begin{subequations} \label{eq:profitss}
\renewcommand{\theequation}{\theparentequation\alph{equation}} 
\begin{alignat}{1}
& \lambda^{(j+1)}_{t,o} = \lambda^{(j)}_{t,o} - \frac{\gamma}{2}r^{(j)}_{t,o}, \quad \forall t, o \label{eq:priceupdate}\\
& r^{(j)}_{t,o} = \sum_{a} \theta^{(j)}_{a,t,o}, \quad \forall t, o  \label{eq:residual} 
\end{alignat}
\end{subequations}


\begin{algorithm}[t]
\caption{ADMM-based Solution}
\textbf{Input:} $(x,\theta_{a,t,o},\lambda_{t,o})$ from centralized risk-neutral solution. \\
\textbf{Output:} Approximate equilibrium of incomplete markets.
\begin{algorithmic}
\State Set $\tau>0$, $J$, $j=0$.
\State Solve $\max_{\Theta_a}\rho_a,\;\forall a\in\mathcal{A}$.
\While{$(Q^{primal}>\tau \lor Q^{dual}>\tau)\land j<J$}
    \State $j\gets j+1$
    \State Update $\lambda_{t,o},\;\pi_{a,o},\;L^{\text{PEN}}_a,\;\forall a$
    \State Solve \eqref{eq:augmentedprob}$\quad\forall a\in\mathcal{A}$
\EndWhile
\end{algorithmic}
\end{algorithm}

Following standard practice in ADMM applications for the optimal exchange problem \cite{Kaminski2021ImpactCompanies, Hoschle2018AnMarkets, Boyd2010DistributedMultipliers}, agents' individual problems are augmented with a term (\ref{eq:pengen}) that penalizes deviations of participants' second-stage decisions between two consecutive iterations. This deviations are weighted with the user-defined parameter $\gamma$ and risk-adjusted probability of each scenario $o$ \cite{Ehrenmann2011GenerationAnalysis}. 
\begin{subequations} \label{eq:penaltytermsection}
\renewcommand{\theequation}{\theparentequation\alph{equation}} 
\begin{alignat}{1}
& L^{\text{PEN}(j+1)}_{a} = \frac{\gamma}{2} \sum_o \sum_t \bigg( W_{t,o} \left( P_o \delta_a + \mu^{(j+1)}_{a,o} \right) \cdot \notag \\
& \quad \quad \quad\quad\quad\quad\quad\quad\bigg( \theta^{(j+1)}_{a,t,o}  - \theta^{(j)}_{a,t,o}  + \bar{r}^{(j)}_{t,o}\bigg)^2 \bigg), \quad \forall a \label{eq:pengen} 
\end{alignat}
\end{subequations}

\noindent where $\bar{r}^{(j)}_{t,o}$ is an average residual, for our case defined as  $r^{(j)}_{t,o}/|A|.$


At each ADMM iteration $(j)$, each agent $a \in \mathcal{A}$ solves the following problem that maximizes its risk-adjusted surplus net of the penalty term:
\begin{align} 
\max_{\Theta_a}\; \rho_a - L^{\text{PEN}(j)}_{a}, \quad 
\text{s.t. } \eqref{eq:gen_cvar1}–\eqref{eq:gen_feas}. \label{eq:augmentedprob}
\end{align}

Our stopping criterion checks the values of primal and dual residuals \cite{Boyd2010DistributedMultipliers}. The primal residual $Q^{P}$ calculates the violation of the market-clearing condition, while the dual residual  $Q^{D}$ tracks the differences in market participants' decisions between consecutive iterations. 
\begin{subequations} \label{eq:primalconvanddual23}
\renewcommand{\theequation}{\theparentequation\alph{equation}} 
\begin{alignat}{1}
& Q^{P} =  \left\| r^{(j)}_{t,o} \right\|_2 
= \sqrt{ \sum_{t} W_{t,o}\sum_{o} P_o\left( r^{(j)}_{t,o} \right)^2 }\label{eq:primal} \\
& Q^{D} = \sum_a \left\| \gamma \bigg(\theta^{(j+1)}_{a,t,o} - \theta^{(j)}_{a,t,o} - \left(\bar{r}^{(j+1)}_{t,o}-\bar{r}^{(j)}_{t,o}\right)\bigg) \right\|_2  \label{eq:dual} 
\end{alignat}
\end{subequations}

The algorithm stops when both residuals fall below a system-dependent tolerance \cite{Boyd2010DistributedMultipliers} defined as
\begin{equation}
  \tau \;=\; \epsilon\,\sqrt{\bigl(|\mathcal{R}|+1\bigr)\,|\mathcal{T}|\,|\mathcal{O}|}\,
  \label{eq:tolerance}
\end{equation} or when the number of iterations reaches 10,000. In the rare event that the maximum number of iterations is reached before the residuals fall below the specified tolerance, the algorithm returns the results from the final iteration.

\section{Case Study}

The simulations are performed for a greenfield investment problem in a copperplate GB-like test system. The case study aims to represent a decarbonized power system that has greenhouse gas emission intensity in line with GB projections for 2040-2050 \cite{seventhcarbonbudget}. The available technologies include combined-cycle gas turbine (CCGT) , CCGT-CCS, nuclear power plant, photovoltaic system (PV), onshore and offshore wind, LDES in the form of hydrogen (H\textsubscript{2}), and a shorter-duration battery energy storage system (BESS), each represented by an individual investor. Investment in nuclear, CCGT, and onshore wind are constrained not to exceed 6\%, 8\%, and 45\% of the peak demand, respectively. Peak demand is scaled to 100MW. Cost data for each technology are summarized in Table~\ref{tab:tablecosts}.


\begin{table}[htbp]
\centering
\caption{Techno-economic Parameters for Each Technology.}
\label{tab:tablecosts}
\resizebox{0.49\textwidth}{!}{%
\begin{tabular}{@{}lcccccccc@{}}
    \toprule
    \makecell{\text{}} &
    \makecell{\text{CCGT}} &
    \makecell{\text{CCGT-CCS}} &
    \makecell{\text{PV}} &
    \makecell{\text{Wind}\\\text{Onshore}} &
    \makecell{\text{Wind}\\\text{Offshore}} &
    \makecell{\text{H\textsubscript{2}}} &
    \makecell{\text{BESS}} &
    \makecell{\text{Nuclear}} \\
    \midrule
    CAPEX [\$/kW] & 1200 & 3000 & 779 & 1315 & 2400 & 2750 & 295 & 6344 \\
    CAPEX [\$/kWh] & - & -& - & - & - & 8 & 156 & - \\
    O\&M [\$/kW-yr] & 29.0 & 85.0 & 15.0 & 31.3 & 89.2 & 88 & 27 & 156 \\
    vO\&M [\$/kWh] & 0.124 & 0.200 & - & 0.006 & 0.001 & - & - & 0.015 \\
    $\eta^{\text{ch}}_s$ [\%] & -& - & - & - & - & 70 & 92 & - \\
    $\eta^{\text{dis}}_s$ [\%] & -& - & - & - & - & 55 & 92 & - \\
    Lifetime [years] & 25 &25& 20 & 30 & 30 & 18 & 20 & 60 \\
    WACC* [\%] & 7.0 &7.0& 6.2 & 6.1 & 6.5 & 7.0 & 6.0 & 7.0\\
    \bottomrule
\end{tabular}%
}
\vspace{-1ex}
\begin{flushleft}
\scriptsize{*Debt fraction: 75.9\%. Tax rate: 25.7\%.}
\end{flushleft}
\end{table}

The uncertainty set includes  15 weather years (scenarios) of time-coincident demand and renewable generation profiles.  To reduce computational complexity, we use varying durations of timesteps $W_{t,o}$ identified based on the clustering method of \cite{salvapineda}. The number of timesteps in each scenario is 672. The flexible demand is set to 10\% of the fixed demand in each scenario and time step. The Value-Of-Lost-Load (VOLL) is defined at \$20300 /MWh to approximate scarcity pricing in the absence of a capacity market \cite{adamsmissingmoney}. Fuel prices for 2040 are assumed to be around \$15.0/MMbtu for natural gas and \$1.2/MMbtu for nuclear fuel. To align with 2040 net-zero objectives, we impose a system-wide emissions-intensity cap of 4 gCO\textsubscript{2}e/kWh (a conservative fraction of the 8 gCO\textsubscript{2}e/kWh reported in \cite{fes2025} to account for unmodeled sources). Unabated gas generation is assumed to emit 330 gCO\textsubscript{2}e/kWh, while CCGT-CCS emits 40 gCO\textsubscript{2}e/kWh.

In this analysis, we vary the weight $\delta$ and keep the CVaR confidence level, $\Psi$, constant at 0.5. All market participants are assumed to have identical risk-aversion coefficients, implying a uniform attitude toward risk. Both the complete and fully incomplete risk market cases are analyzed under these assumptions, with the corresponding results presented in the following section.

\section{Results}

\subsection{Investment under Complete Risk Markets}
\label{iva}
\begin{figure}[htb!]
    \centering
    \includegraphics[width=\linewidth]{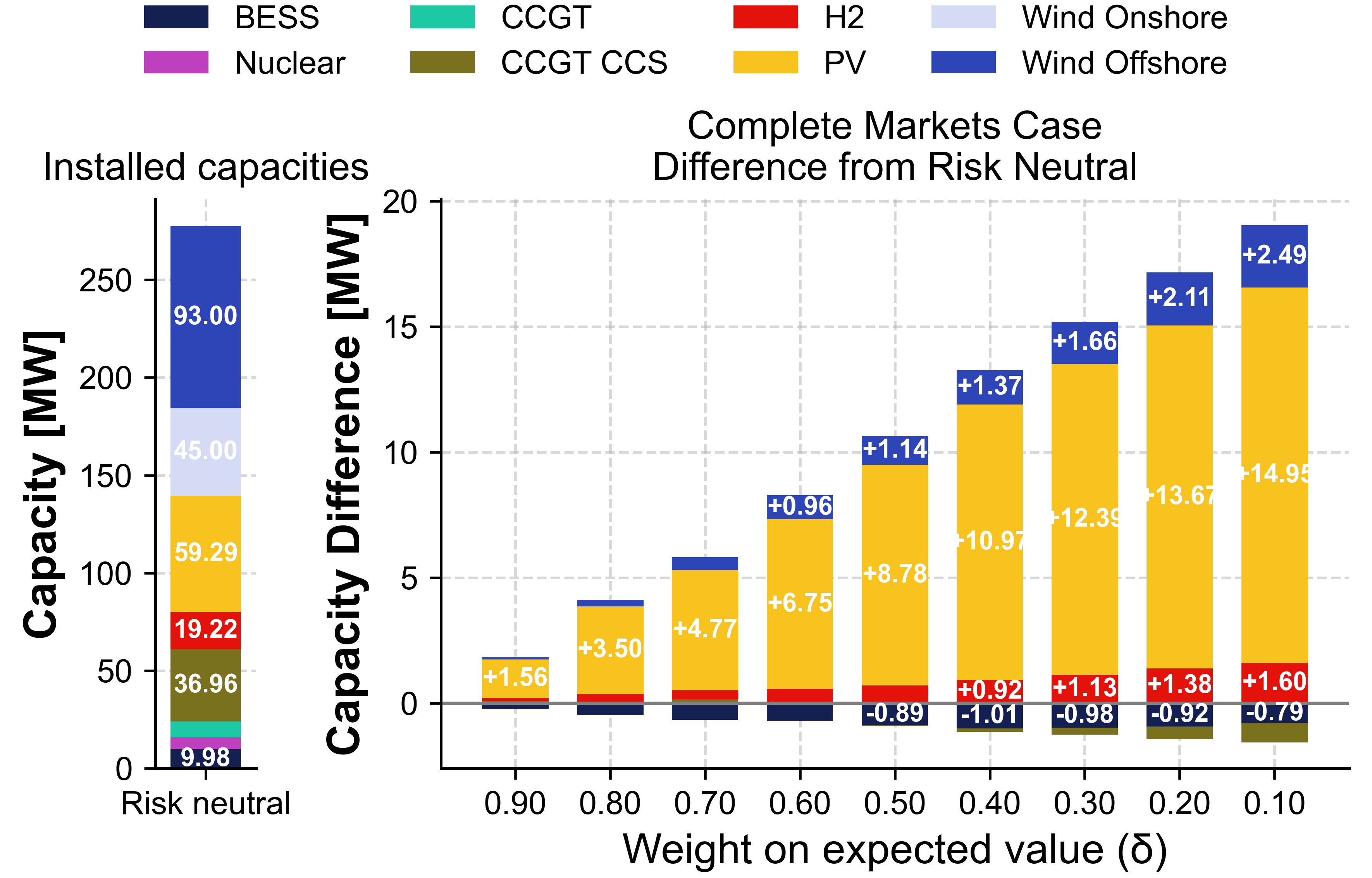}
    \caption{Left: Installed capacity by technology in the risk-neutral case. Right: Investments under complete risk markets across mean–CVaR weights ($\delta$), reported as change in installed capacity by technology relative to the risk-neutral case. The storage duration is approximately 15 and 116 hours for BESS and LDES, respectively.}
    \label{fig:installed_capacities1}
\end{figure}

We first analyse the investments for a case, hereafter called \textit{complete risk markets} as in section \ref{SUB: CompleteMarketsForm}, where risk-averse investors have access to any hedging instruments that could help them manage financial risks. We present the capacity mix for the complete risk markets case, under different risk aversion levels in Fig.~\ref{fig:installed_capacities1}. 

The generating capacity is predominantly from clean energy sources, with nuclear, CCGT-CCS, and VRE contributing 234 to 251 GW of power capacity. Storage power capacity is split between long-duration hydrogen energy storage (19–21 MW) and BESS 9-10 MW).

As risk aversion increases (i.e., as $\delta$ decreases), total power capacity rises in order to better capture welfare under tail scenarios. This increase is driven primarily by capacity from solar PV and offshore wind, complemented by LDES. Meanwhile, the capacities of CCGT-CCS and BESS are scaled back. The capacities for nuclear and offshore wind remain at their maximum allowable levels across all cases. Overall, the shift in investments raises system reliability, with the expected unserved energy (EUE) reducing by 35\% between $\delta=1$ and $\delta=0.1$. Simultaneously, the average price level falls by approximately 17\% over that same $\delta$ range.

\subsection{Effects of Missing Risk Markets}
\begin{figure}[htb!]
    \centering
    \includegraphics[width=1.0\linewidth]{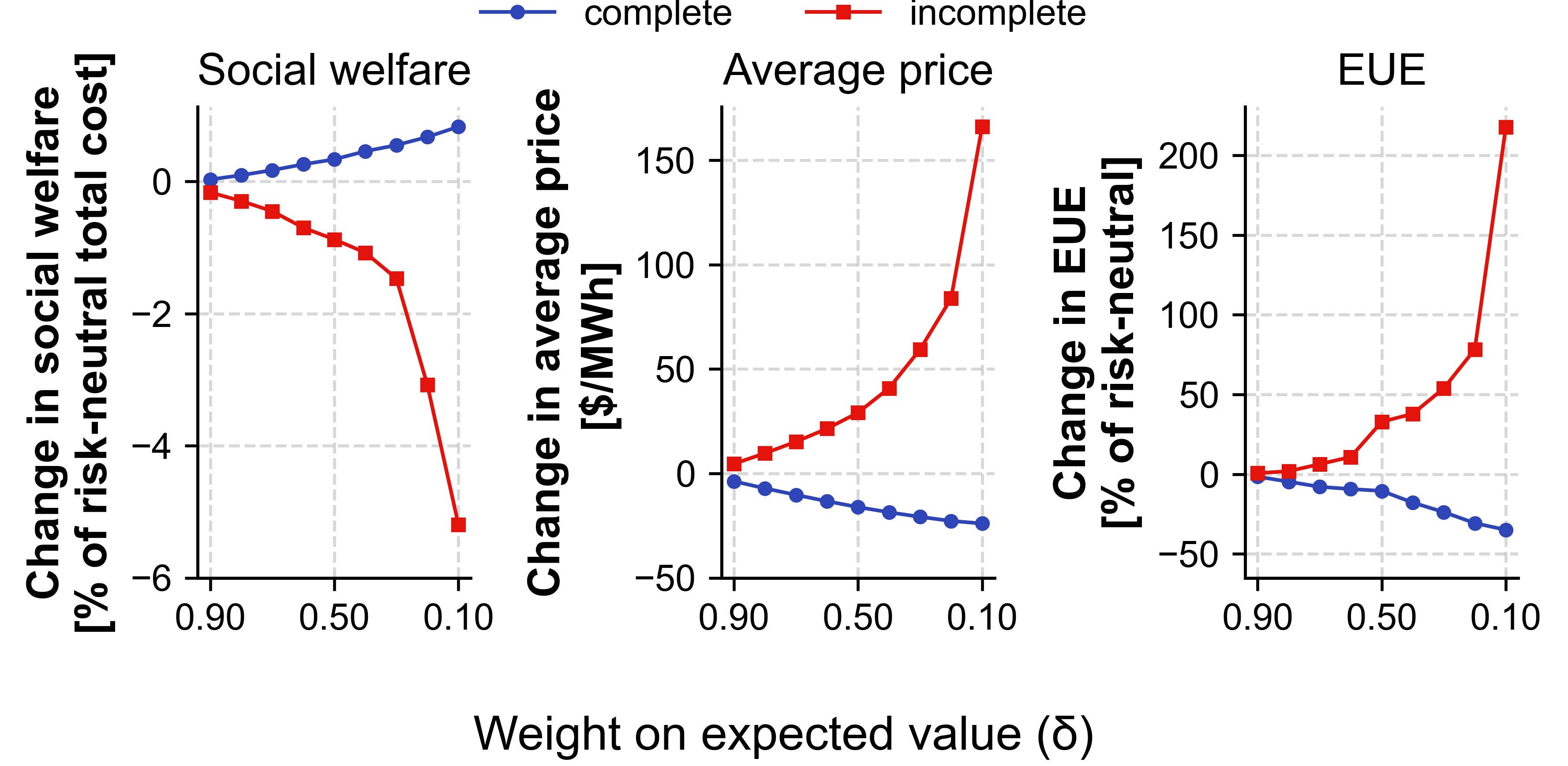}
    \caption{Changes in social welfare* between complete and fully incomplete cases (\% of risk-neutral total system costs) and in average energy price and EUE relative to the risk-neutral case.}
    \label{fig:threepanel}
\begin{flushleft}
\scriptsize{*The reported change reflects the risk-adjusted social welfare.}
\end{flushleft}
\end{figure}

As expected from theory, missing risk markets negatively affect the social welfare (see left panel in Fig.~\ref{fig:threepanel}). The social welfare consists of both producers' and consumers' surplus. Under incomplete markets, consumers pay on average higher energy prices for worse reliability levels (see middle and right panels in Fig.~\ref{fig:threepanel}). The more risk-averse investors and consumers are, the greater the adverse impact. These negative outcomes arise because consumers cannot establish contractual agreements with investors. Instead, they reveal their preferences solely through their electricity consumption decisions. As a result, even though consumers’ underlying willingness to pay is unchanged, the inability to engage in risk trading prevents these preferences from being signaled through contract pricing. This mismatch leads to underinvestment in capacity (from the consumer’s perspective), which in turn produces higher prices and lower welfare.

\renewcommand{\arraystretch}{1.4} 

\subsection{Investment under Fully Incomplete Risk Markets}
\begin{figure}[htb!]
    \centering
    \includegraphics[width=\linewidth]{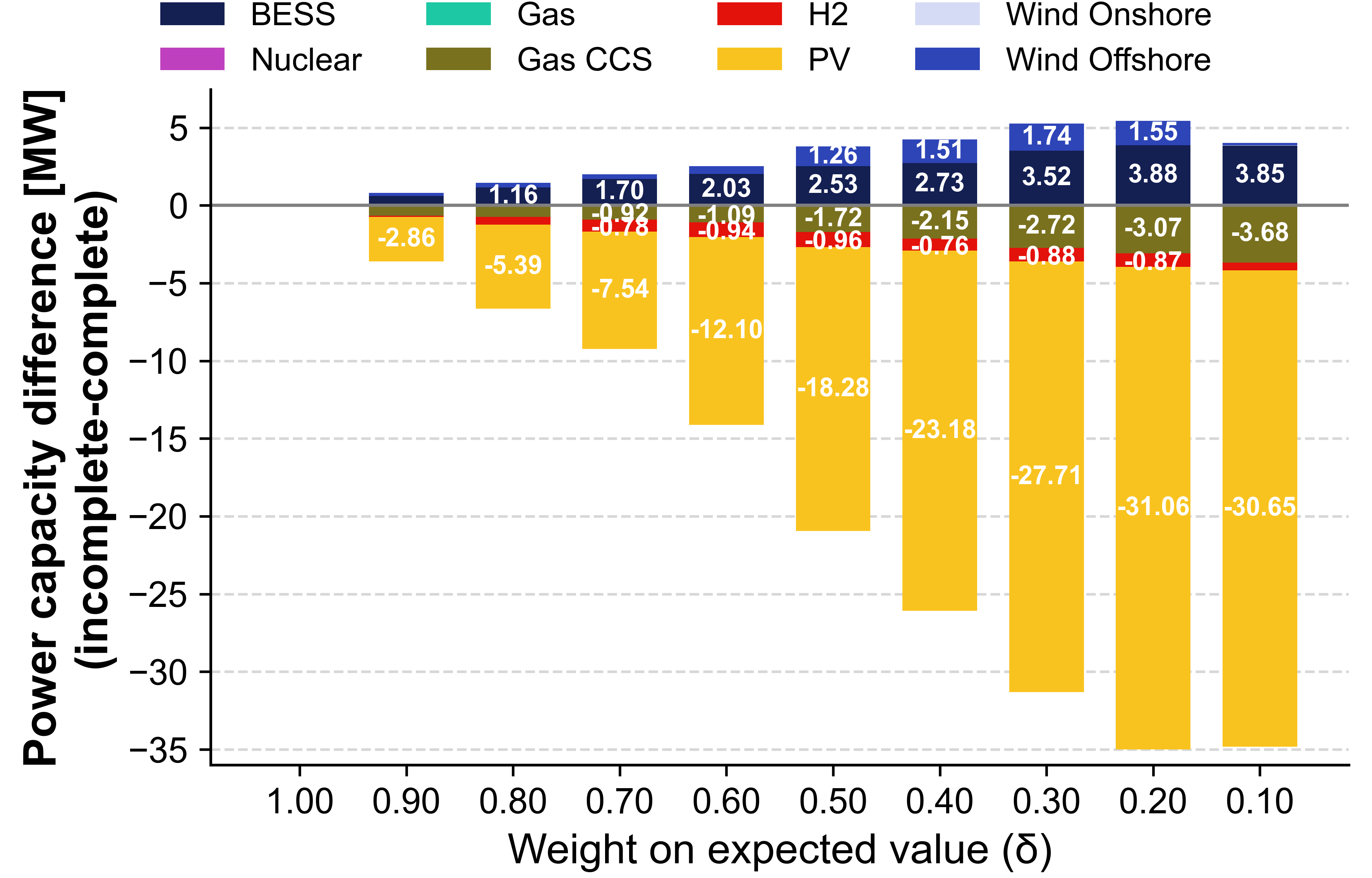}
    \caption{Investments under incomplete risk markets across mean–CVaR weights($\delta$), reported as change in installed capacity by technology relative to the complete risk markets case.}    \label{fig:installed_capacitiesincomplete}
\end{figure}

To evaluate the impact of missing risk markets on investments, we contrast the results of a fully incomplete case with the results presented in section \ref{iva}. Fig.~\ref{fig:installed_capacitiesincomplete} presents the capacity differences between the two cases. Overall, total capacity investments decline in the fully incomplete markets case, and the technology mix changes. Specifically, investments in PV, LDES, and CCGT-CCS decrease, while there is an increase in BESS and offshore wind. 

\begin{figure}[htb!]
    \centering
    \includegraphics[width=\linewidth]{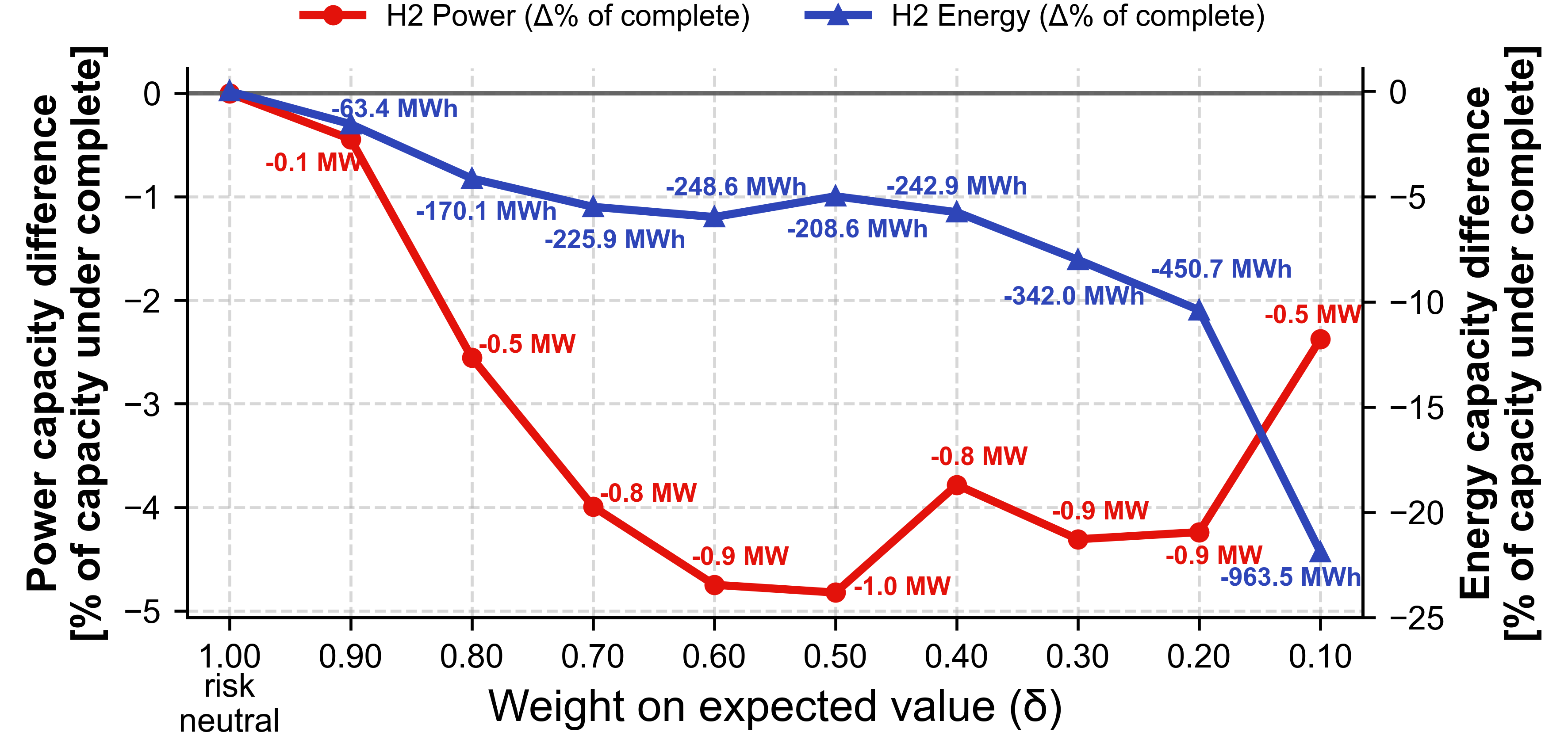}
    \caption{Percentage change in LDES power and energy capacity investments (incomplete–complete) across mean–CVaR weights($\delta$). Labels show the associated changes in MW/MWh.}    \label{fig:ldes_storage}
\end{figure}

For LDES, whereas the change in its power capacity appears relatively small (less than 5\%), the change in the energy capacity of LDES is substantial (up to 20\% for $\delta{=}0.1$) (see Fig.~\ref{fig:ldes_storage}). These results show that the lack of risk markets will result in more severe under-investment (compared to the complete market case) for the energy rather than the power component of LDES. In other words, the lack of risk markets results in investments with shorter durations of LDES. For instance, at $\delta=0.1$, the duration of LDES drops from 116 hours to 93 hours when the hedging contracts are missing. 

These results arise because investors in a given technology cannot establish contractual agreements with other technology providers or consumers. Consequently, while consumers would benefit from improved reliability due to additional investments in LDES and CCGT-CCS, investors in those technologies are more concerned about potential tail losses than motivated by the tail profits. Our findings are consistent with previous research, which has also shown that in fully incomplete markets, investments in reliability-enhancing technologies with high capital costs tend to decline, and technologies with lower capital costs are preferred \cite{Mays2019AsymmetricMarkets, Dimanchev2024ConsequencesEmissions}.

\subsection{Impact of Missing Risk Markets on Storage Financing Costs}


In a risk-averse environment, the cost of capital increases with the volatility of investment returns. Using the formula presented in Mays and Jenkins \cite{Mays2023FinancialPenetration}, we calculate $\text{WACC}^{\text{IMP}}_r$, the cost of capital inclusive of a risk premium, as follows:
\begin{equation}
\begin{aligned}
    & \sum_{l=1}^{L_r} \frac{\bar{C}^{\text{REV}}_{r}}{(1 + \text{WACC}^{\text{IMP}}_r)^l} = \sum_{l=1}^{L_r} \frac{C^{\text{INV}}_{r}}{(1 + \text{WACC}_r)^l} \label{eq:impliedwacc1}
\end{aligned}
\end{equation}

where $L_r$ represents the lifetime of resource $r$, $\text{WACC}_r$ is the risk-free weighted average cost of capital, $\bar{C}^{\text{REV}}_{r}$ is the average net revenues of the resources, weighted by the probability of a weather year. 

\begin{figure}[t]
    \centering
    \includegraphics[width=\linewidth]{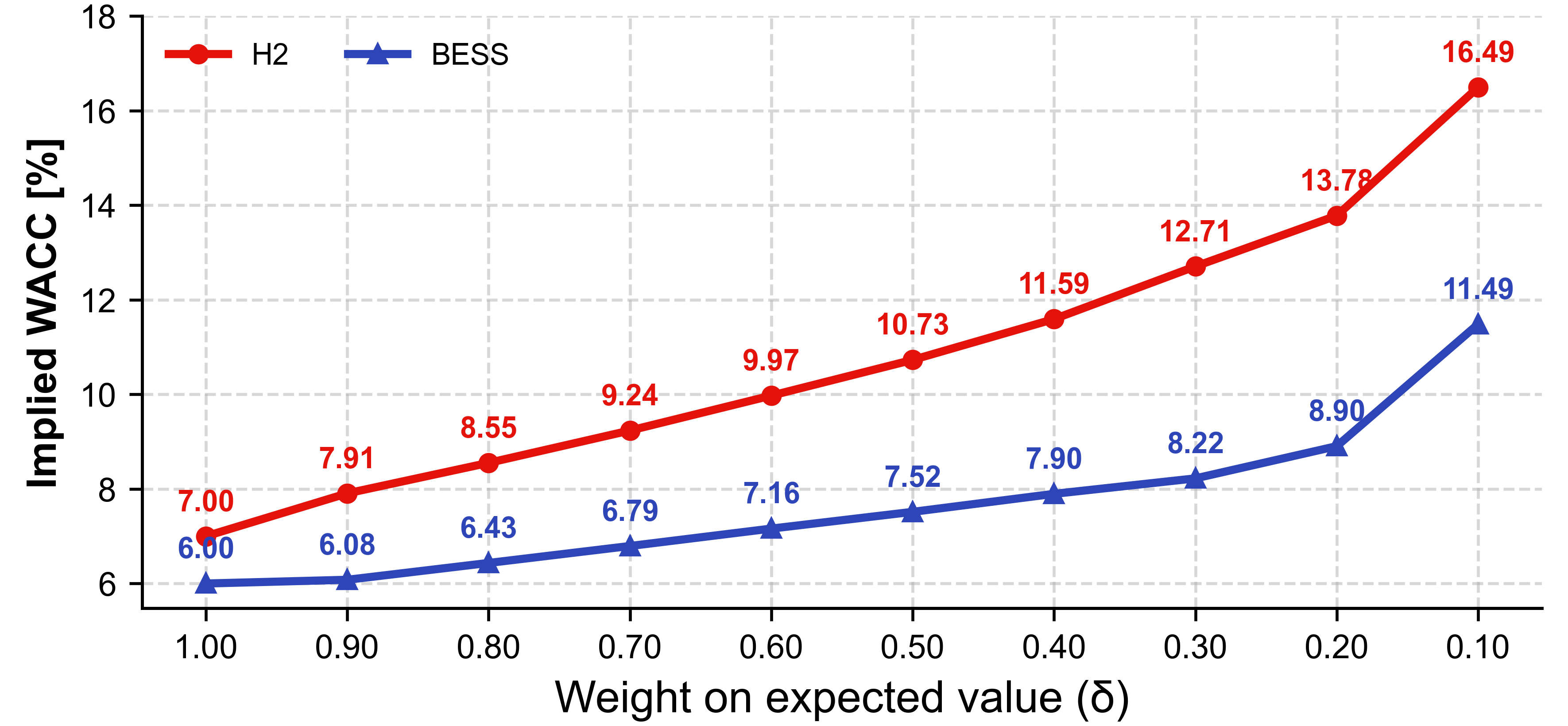}
    \caption{Effect of fully incomplete markets on WACC under varying levels of risk aversion for storage assets.}    \label{fig:wacc}
\end{figure}

Fig.~\ref{fig:wacc} displays the $\text{WACC}^{IMP}_{r}$, calculated using (\ref{eq:impliedwacc1}), for BESS and LDES in fully incomplete markets. The implied cost of capital increases with the degree of risk aversion and is higher for LDES than for BESS, since the higher round-trip efficiency of BESS results in less volatile net revenues for BESS. The risk premium, i.e., the difference between the implied cost of capital and the risk-free cost of capital, ranges from 0.91\% to 9.49\% for LDES, which is substantially higher than the corresponding 0.08\%–5.49\% range for BESS. These findings affirm the critical role of risk management in enabling LDES investments, highlighting the potential influence of policy mechanisms aimed at de-risking LDES investments.

\section{Conclusions}
This article investigates how missing risk markets influence investment decisions in a deeply decarbonized power system. We formulate the capacity expansion problem as a two-stage stochastic equilibrium model and solve it using the ADMM algorithm. The methodology is applied, for the first time, to a problem that independently sizes the power and energy components of storage assets. Using a case study of Great Britain with multi-year VRE uncertainty, we provide novel insights into the effects of incomplete risk markets on clean energy technologies, including long-duration energy storage.

The results show that missing risk markets reduce social welfare and lead to adverse outcomes for consumers, who face higher electricity prices and lower reliability levels. The absence of risk-hedging opportunities discourages investment in technologies with high capital costs, such as long-duration energy storage, that enhance reliability but face substantial inter-annual revenue volatility. Under-investment disproportionately affects the energy rather than the power capacity of long-duration energy storage. The cost of capital for long-duration energy storage includes a higher risk premium relative to shorter-duration storage technologies. This risk premium, ranging from approximately 1\% to 9\% in our results, illustrates how limited revenue certainty can hinder investment in long-duration energy storage and underscores the potential importance of de-risking mechanisms to accelerate its deployment.

Future work could apply the proposed methodological framework to case studies with a more detailed power system representation, including network constraints and cross-border interconnections. Further extensions could incorporate additional market products, such as reserves and redispatch services, which provide supplementary revenue streams for storage. Methodologically, the framework could be expanded to capture heterogeneous risk attitudes, participants with diversified technology portfolios, and policies and other contracts that could hedge investors' revenues.



%

\section*{Data and Code Availability}

The data, code and other supporting materials used to generate the results of this study are available on GitHub at: \url{https://github.com/Andreas-Makrides-ICL/incomplete_markets_long_duration_energy_storage}

\bibliographystyle{IEEEtran}
\bibliography{references} 
%


\end{document}